\documentclass[showpacs,twocolumn,prc,preprintnumbers,amsmath,amssymb]{revtex4-1}
\usepackage{graphicx}
\usepackage{dcolumn}
\usepackage{bm}
\usepackage{color}
\usepackage[usenames,dvipsnames]{xcolor}

\begin{document}
 
\title{Effects of initial state fluctuations in the final state elliptic flow measurements using the NeXSPheRIO model}

\author{R.~Derradi de Souza}\email{rderradi@ifi.unicamp.br}
\author{J.~Takahashi}
\affiliation{Universidade Estadual de Campinas, S\~{a}o Paulo, Brazil}

\author{T.~Kodama}
\affiliation{Universidade Federal do Rio de Janeiro, Rio de Janeiro, Brazil}

\author{P.~Sorensen}
\affiliation{Brookhaven National Laboratory, New York, USA}


\begin{abstract}
We present a systematic study of the effects due to initial condition fluctuations in systems formed by heavy-ion collisions using the hydrodynamical simulation code NeXSPheRIO.
The study was based on a sample of events generated simulating Au+Au collisions at center of mass energy of 200 GeV per nucleon pair with impact parameter ranging from most central to peripheral collisions.
The capability of the NeXSPheRIO code to control and save the initial condition (IC) as well as the final state particles after the 3D hydrodynamical evolution allows for the investigation of the sensitivity of the experimental observables to the characteristics of the early IC.
Comparisons of results from simulated events generated using fluctuating initial conditions and smooth initial condition are presented for the experimental observable elliptic flow parameter ($v_2$) as a function of the transverse momentum, $p_t$, and centrality.
We compare $v_2$ values estimated using different methods, and how each method responds to effects of fluctuations in the initial condition.
Finally, we quantify the flow fluctuations and compare to the fluctuations of the initial eccentricity of the energy density distribution in the transverse plane.
\end{abstract}

\keywords{elliptic flow, hydrodynamic model, initial state fluctuations, heavy ion collisions}

\pacs{25.75.Ld, 24.10.Nz, 24.60.Ky, 25.75.-q}

\maketitle
\setcounter{page}{1}

\section{Introduction} \label{sec:introduction}
In many dynamical systems, the evolution of the system affects its physical characteristics and the information of the early stages is lost, when observing only the probes from the final stages.
However, by studying how much of the initial condition information is retained and survives the dynamical evolution it is possible to obtain valuable insight not only on the characteristics of the initial conditions but also on the interactions that occur in the evolution.
These conditions can be applied in many physical systems such as in the study of the evolution of our universe and the anisotropy observed in the CMB \cite{ApJ.142.419,*arXiv:astro-ph.9511130,*arXiv:astro-ph.0305591}, or in the study of the nuclear reactions in high energy \cite{PhysRevC.77.064902}.

In relativistic heavy-ion collisions an important experimental observable is known as the elliptic flow.
Due to geometrical anisotropy created in the initial condition by the two colliding nuclei and the evolution of the system, a space-momentum correlation develops and an anisotropy in the azimuthal distribution of the final particles can be measured.
The elliptic flow or the Fourier expansion second harmonic ($v_2$) corresponds to the amplitude of this azimuthal anisotropy.
Hydrodynamic models predict that indeed elliptic flow should be sensitive to the eccentricity of the initial conditions \cite{PhysRevD.46.229}.
Model calculations of $v_2$ are also in good agreement with much of the experimental data \cite{NuclPhys.A696.197,PhysLett.B503.58,PhysLett.B583.73,NuclPhys.A757.102,NuclPhys.A757.184,PhysRevLett.99.172301,PhysRevLett.106.192301,PhysRevLett.106.042301,PhysRevC.82.014903}.
Furthermore, two or more particle correlations may carry important information on the initial state, hence on the mechanism of QCD dynamics at the very early stage of the collision.
Recent studies \cite{PhysRevLett.93.182301,PhysRevLett.97.202302,BrazJPhys.37.717,BrazJPhys.37.99,PhysRevLett.101.112301,NuclPhys.A698.639,arXiv:nucl-ex.0312008,PhysRevLett.103.242301,PhysRevC.72.064911,JPhysG.38.045102} show that a scenario including fluctuations of the initial condition in addition to the geometrical eccentricity describes better the experimental data than without the inclusion of initial state fluctuations. 
These fluctuations in the initial condition will affect the eccentricity by changing the average value and also by introducing an additional fluctuation which will then affect the final observed $v_2$ values.
To explore the effects of these fluctuations, we need a more precise study, quantifying the effects of these initial condition fluctuations to the final experimental observables.
For this purpose we have used the simulation code NeXSPheRIO \cite{PhysRevC.65.054902,*NuclPhys.A698.639,*BrazJPhys.35.24} that allows the generation of events using smooth or fluctuating initial conditions (IC) and obtain the final particles after the hydrodynamical evolution.
The procedures used in experimental data analysis for $v_2$ determination are applied to the simulated data and the effects of the initial fluctuations are investigated through the differences in the $v_2$ estimates using different methods.
NeXSPheRIO allows for the control and study of both the initial condition before the hydrodynamical evolution and the final particles after hadronization and freeze-out.
Thus, it allows for a detailed study, on an event-by-event analysis, of the correlation between initial condition parameters such as eccentricity and fluctuations to the final state observables such as $v_2$.
This simulation code has already been extensively tested and has presented reasonable agreement with experimental data \cite{BrazJPhys.37.717,arXiv:0711.4544,JPhysG.31.S1041}.
In the next section we describe how the initial energy density is obtained in the NeXSPheRIO code and how the eccentricity is calculated. In section \ref{sec:ellipticFlowDetermination} we describe the methods used to estimate the elliptic flow. In section \ref{sec:results} we present our results and, finally, in section \ref{sec:summary} we summarize our conclusions.

\section{Initial energy density} \label{sec:initialEnergyDensity}
The NeXSPheRIO code allows the use of smooth and fluctuating initial conditions, which is particularly suited and convenient to study the effects of fluctuations in the elliptic flow calculations.
The initial conditions are generated by a microscopic model called NeXuS \cite{PhysRevC.65.054902} which produces, on an event-by-event basis, detailed space distributions of the energy-momentum tensor, baryon-number, strangeness and charge densities \cite{BrazJPhys.35.24}.
The initial condition is then used as input to the hydrodynamical model and the system evolution is computed up to a given point where a decoupling mechanism is applied and final particles are produced \cite{BrazJPhys.35.24}.
Considering the transverse profile of the initial energy density distribution around $\eta=0$, it is possible to calculate the eccentricity of the initial geometry.
In the situation where there is no fluctuation and the colliding nuclei are considered to have smooth distributions, the major axis of the almond shape of the overlap area at the moment of the collision is perpendicular to the plane defined by the impact parameter and the beam axis, generally referred to as reaction plane, while the minor axis coincides with the direction of the impact parameter.

\begin{figure}[ht]
\includegraphics[width=8.50cm]{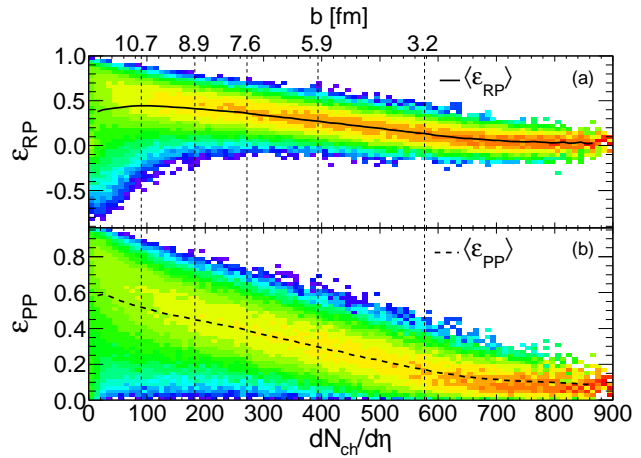}
\caption{\label{fig:EccentVsdNchdeta} (color online) (a) Reaction plane eccentricity $\varepsilon_{RP}$ and (b) participant plane eccentricity $\varepsilon_{PP}$ as a function of the number of charged particle produced at $\eta = 0$. The black line in panel (a) represents the mean of the $\varepsilon_{RP}$ distribution, which is equal to the $\varepsilon_{RP}$ and $\varepsilon_{PP}$ of the smooth IC case. The dashed line in the panel (b) represents the average of the $\varepsilon_{PP}$ for the fluctuating IC case.}
\end{figure}

The eccentricity calculated with respect to this reaction plane, defined by $\varepsilon_{RP}$, is given by \cite{PhysRevC.77.014906}:
\begin{equation}
\varepsilon_{RP}=\frac{\sigma_y^2-\sigma_x^2}{\sigma_y^2+\sigma_x^2},
\end{equation}
where $\sigma_x^2$ and $\sigma_y^2$ are the variances of the distribution along $x$ and $y$ directions, respectively.
Because of fluctuations, the actual distribution of the hot material created by collisions of participants can have the principal axis different from those of the smooth initial condition and the minor axis deviates from the reaction plane direction defined above \cite{PhysRevC.77.014906}.
The plane that maximizes the eccentricity determined from the initial energy distribution is generally called participant plane \cite{arXiv:0905.0174}.
The participant plane eccentricity $\varepsilon_{PP}$ is given by \cite{PhysRevC.77.014906}:
\begin{equation}
\varepsilon_{PP}=\frac{\sqrt{\left( \sigma_y^2 - \sigma_x^2 \right)^2 + 4 \sigma_{xy}^2}}{\sigma_x^2 + \sigma_y^2},
\end{equation}
where $\sigma_{xy}$ is the covariance of the transverse energy density distribution.
Note that the reaction plane is determined just by the initial geometry of the collision for the whole system uniquely, whereas the participant plane depends on the dynamics of the collisions in each event.
In Fig. \ref{fig:EccentVsdNchdeta}, we show the distributions of the reaction plane eccentricity and the participant plane eccentricity as a function of the collision impact parameter. In the x-axis of Fig. \ref{fig:EccentVsdNchdeta}, we also show the equivalent number of charged particle density produced at $\eta = 0$, $dN_{ch}/d\eta$, which is obtained after the full event simulation procedure is completed.
The top panel shows the event-by-event reaction plane eccentricity $\varepsilon_{RP}$ and the bottom panel shows the participant plane eccentricity $\varepsilon_{PP}$.
By construction, the smooth IC reaction plane eccentricity corresponds to the mean eccentricity of the fluctuating IC case, represented by the solid black line in the panel (a) of Fig. \ref{fig:EccentVsdNchdeta}. Also, the participant plane eccentricity matches the reaction plane eccentricity for the smooth IC case.
For the most peripheral collisions the reaction plane eccentricity can assume negative values due to fluctuation that will force the asymmetry plane to be perpendicular to the reaction plane.
The dashed line in the panel (b) represents the mean of the participant plane eccentricity for the fluctuating IC case only.

\section{Elliptic flow determination} \label{sec:ellipticFlowDetermination}
The momentum anisotropy of the final particles azimuthal distributions is generally described by a Fourier expansion of the azimuthal angle $\phi$ of each particle with respect to the reaction plane angle $\Psi_{RP}$ \cite{PhysRevC.58.1671}.
The most relevant contribution comes from the second coefficient of the expansion, $v_2$, which is due to the almond shape of the initial overlap area of the incident nuclei in the transverse plane.
However, the reaction plane angle cannot be directly measured experimentally and, therefore, several methods and techniques have been developed to estimate the $v_2$ coefficient from the experimental data \cite{PhysRevC.58.1671,arXiv:0809.2949,PhysRevC.64.054901,PhysRevC.83.044913}.
For this work, we have computed the elliptic flow coefficients in several different ways: (a) using the true reaction plane ($v_2\{\mathrm{RP}\}$) known in the simulation; (b) estimating a reference plane with the event plane method ($v_2\{\mathrm{EP}\}$) \cite{arXiv:0809.2949,PhysRevC.58.1671}; (c) calculating the participant plane from the initial energy density distribution profile around $\eta=0$; and (d) using the formalism of the cumulant method ($v_2\{2\}$ and $v_2\{4\}$) \cite{PhysRevC.83.044913}.
Below we summarize how each method works.

The reaction plane elliptic flow, $v_2\{\mathrm{RP}\}$, is calculated with respect to the true reaction plane as:
\begin{equation}
v_n\{\mathrm{RP}\} = \langle \cos \left[ n \left( \phi - \Psi_{RP} \right) \right] \rangle,
\end{equation}
where $\phi$ is the azimuthal angle of each particle, $\Psi_{RP}$ is the true reaction plane angle and the angle brackets denotes averages first over all selected particles in each event, and then over all events.
In NeXSPheRIO, for fluctuating IC, an event is first generated by generating nucleons inside the colliding nuclei.
Thus, before the collision, the colliding matter already looses its symmetry with respect to the reaction plane, as defined by the impact parameter vector and the beam axis. 
In this sense, the concept of reaction plane looses its physical meaning for event-by-event calculations.
However, we keep the calculations of $v_2\{\mathrm{RP}\}$ as a reference for comparisons with the flow results obtained with other methods.

\begin{figure}[htb]
\includegraphics[width=8.50cm]{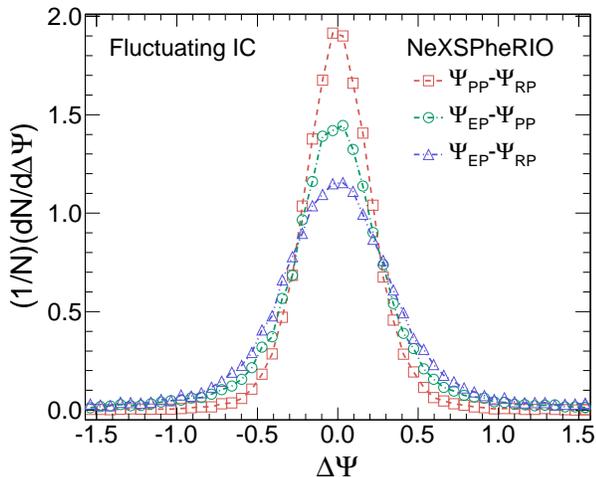}
\caption{\label{fig:PsiCorr} (color online) Correlation between different reference plane angles for mid-central NeXSPheRIO events. Red open squares (dashed line) show the difference between $\Psi_{PP}$ and $\Psi_{RP}$, green open circles (dot-dashed line) show the difference between $\Psi_{EP}$ and $\Psi_{PP}$, and blue open triangles (double-dot-dashed line) show the difference between $\Psi_{EP}$ and $\Psi_{RP}$.}
\end{figure}

\begin{figure}[htb]
\includegraphics[width=8.50cm]{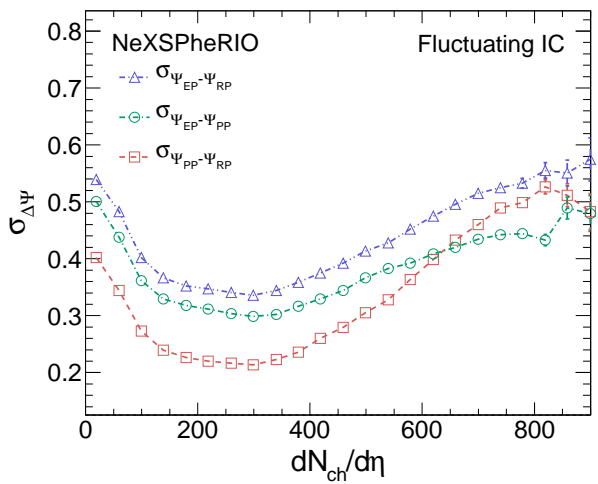}
\caption{\label{fig:sigPsiCorr} (color online) Width of the correlation between different reference plane angles as a function of the number of charged particles at $\eta = 0$. Red open squares (dashed line) show the width of $\Psi_{PP} - \Psi_{RP}$, green open circles (dot-dashed line) show the width of $\Psi_{EP} - \Psi_{PP}$, and blue open triangles (double-dot-dashed line) show the width of $\Psi_{EP} - \Psi_{RP}$.}
\end{figure}

The event plane method is used to estimate the true reaction plane using the flow anisotropy itself.
It is calculated by defining the flow vector as follows \cite{arXiv:0809.2949}:
\begin{equation}
\Psi_n = \frac{\tan^{-1}\left( Y_n / X_n  \right)}{n}, \quad
\left \{ \begin{array}{rl}
X_n = \sum_i w_i \cos \left( n \phi_i  \right) \\
Y_n = \sum_i w_i \sin \left( n \phi_i  \right)
\end{array} \right.
\end{equation}
where $w_i$ is a weight and $\phi_i$ is the azimuthal angle of particle $i$.
The flow coefficients are then determined by \cite{arXiv:0809.2949}:
\begin{equation}
v_n\{\mathrm{EP}\} = \frac{\langle \cos [ n \left( \phi - \Psi_n  \right) ] \rangle}{\langle \cos \left[ n \left( \Psi_n - \Psi_{RP} \right) \right] \rangle},
\label{eq:v2ep}
\end{equation}
where the denominator in the right-hand side of the equation \ref{eq:v2ep} is the resolution of the event plane, and the angle brackets denote averages taken over all selected particles in each event, and then over all events.
Experimentally, $\Psi_{RP}$ is not known and thus the resolution must be inferred in some other way.
The technique applied here and extensively used in experimental data analysis consists in dividing the event being analyzed into two sub-events (event \textit{a} and event \textit{b}) of same multiplicity and calculate the resolution of the sub-event.
It has been shown that the resolution of the sub-events is given by \cite{PhysRevC.58.1671}:
\begin{eqnarray}
\langle \cos \left[ n \left( \Psi_n^a - \Psi_{RP} \right) \right] \rangle = \langle \cos \left[ n \left( \Psi_n^b - \Psi_{RP} \right) \right] \rangle \nonumber \\ 
= \sqrt{ \langle \cos \left[ n \left( \Psi_n^a - \Psi_n^b \right) \right] \rangle},
\end{eqnarray}
where $\Psi_n^a$ and $\Psi_n^b$ are the event plane angles for the sub-events \textit{a} and \textit{b}, respectively.
From the resolution of the sub-events one is able to estimate the resolution of the event plane for the full event (see details in Refs. \cite{PhysRevC.58.1671,arXiv:0809.2949}).
The event plane angle can deviate from the true reaction plane angle due to the fluctuations in the initial conditions and also due to the resolution caused by the limited statistics of the measured particles.
Neglecting the angle shift due to the statistical resolution, it is expected that $\Psi_{EP}$ would follow $\Psi_{PP}$, the participant plane azimuthal angle. So, in order to check this we also calculate $v_2$ with respect to the participant plane angle: 
\begin{equation}
v_n\{\mathrm{PP}\} = \langle \cos \left( n \left[ \phi - \Psi_{PP} \right] \right) \rangle,
\label{eq:v2pp}
\end{equation}
where $\Psi_{PP}$ is calculated from the initial energy density transverse distribution profile as \cite{PhysRevC.77.014906}:
\begin{eqnarray}
\Psi_{PP} &=& \tan^{-1}\left( \frac{\pm\sigma_{xy}}{\sigma_y^2 - \lambda^{\mp}}  \right), \\
\lambda^{\pm} &=& \frac{1}{2} \left( \sigma_y^2 + \sigma_x^2 \pm \sqrt{\left( \sigma_y^2 - \sigma_x^2 \right)^2 + 4 \sigma_{xy}}  \right).
\end{eqnarray}
A recent study \cite{PhysRevC.83.034901} showed that the event plane angle $\Psi_{EP}$ as estimated from the azimuthal particle distribution itself seems to be more correlated to the participant plane angle $\Psi_{PP}$ than to the reaction plane angle $\Psi_{RP}$.
Following this study we have also computed the distributions of the correlations between $\Psi_{RP}$, $\Psi_{EP}$ and $\Psi_{PP}$. In Fig. \ref{fig:PsiCorr} we show the distribution of the difference between the different angles, calculated for NeXSPheRIO mid-central events.
The difference $\Psi_{PP}-\Psi_{RP}$ (shown in red open squares symbols) provides the magnitude of the variation in the participant plane direction caused by event-by-event fluctuations in the initial energy density profile.
The distribution of $\Psi_{EP}-\Psi_{RP}$ (shown in blue open triangles) represents the magnitude of the event plane dispersion caused by the IC fluctuations convoluted to the effect of resolution caused by the limited number of particles used in the $\Psi_{EP}$ determination.
The distribution of $\Psi_{EP}-\Psi_{PP}$ (shown in open green circles) is formed by the statistical resolution and the width of the true correlation between $\Psi_{EP}$ and $\Psi_{PP}$, where by true correlation we mean the part that comes from the initial state fluctuation.
For instance, assuming that the participant plane and the event plane are totally correlated (as in an event with smooth IC), the distribution of the difference between these two angles would be due to the statistical resolution only.
The width ($\sigma$) of the distribution of the angle difference as a function of the collision centrality is summarized in Fig. \ref{fig:sigPsiCorr}.
From this plot we can see a decrease of the angular resolution for the very peripheral collisions and also for the very central collisions.
The increase of the width for central collision is caused by the decrease of the initial eccentricity as well as the increase of the lumpiness in the initial energy density distribution, therefore, enhancing the effect of fluctuations.

An alternative method used in experimental data analysis to estimate the elliptic flow is through the cumulant formalism.
The cumulant method calculates the flow coefficients directly from particle correlations, without the explicit need of a reference plane.
The prescription used in this work is called Q-cumulants or direct cumulants, presented in Ref. \cite{PhysRevC.83.044913}.
The procedure is divided into two parts where in the first part, the reference flow is calculated using all particles inside the selected range. In the second part, the differential flow ($p_t$ dependent) is then calculated for the particles of interest.
By following the notation used in Ref. \cite{PhysRevC.83.044913}, the flow coefficients for second and fourth order cumulants can be written as:
\begin{eqnarray}
\textrm{Reference flow: } \left \{ \begin{array}{l}
v_n\{2\} = \sqrt{\langle \langle 2 \rangle \rangle} \\ 
v_n\{4\} = \sqrt[4]{-\left[ \langle \langle 4 \rangle \rangle - 2 \cdot \langle \langle 2 \rangle \rangle^2 \right]},
\end{array} \right. \\ \nonumber \\
\textrm{Differential flow: } \left \{ \begin{array}{l}
v'_n\{2\} = \frac{\langle \langle 2' \rangle \rangle}{\sqrt{\langle \langle 2 \rangle \rangle}} \\
v'_n\{4\} = \frac{- \left[ \langle \langle 4' \rangle \rangle - 2 \cdot \langle \langle 2' \rangle \rangle \langle \langle 2 \rangle \rangle \right]}{\sqrt[4]{- \left[ \langle \langle 4 \rangle \rangle - 2 \cdot \langle \langle 2 \rangle \rangle^2  \right]^3}},
\end{array} \right.
\end{eqnarray}
where the double brackets denote weighted averages of 2- and 4-particle correlations, first over the particles and then over the events.
The weights are the total number of combinations from two or four particle correlations, respectively, and they are used to minimize the effects due to multiplicity fluctuations. The advantage of the Q-cumulant method is that it is not necessary to perform nested loops to compute all possible combinations in multi-particle correlations.
Instead, it uses the flow vectors to calculate directly the multi-particle cumulants (see details in Ref. \cite{PhysRevC.83.044913}).
The derivation of the expressions for higher order cumulants is straightforward.
We present here only up to the fourth order since the sixth and higher orders do not seem to differ much from the fourth (see for instance Ref. \cite{PhysRevC.72.014904}).
In the next section we present the details of the simulated sample used, the centrality classes, and the estimates for $v_2$ obtained using each one of the methods described above.

\section{Results} \label{sec:results}
In order to compare the elliptic flow estimates obtained with different techniques, we used a sample of simulated events for Au+Au collisions at the center of mass energy of 200 GeV per nucleon pair.
For the analysis presented here we used only the charged particles produced and the weak decays have been turned off to minimize the non-flow contribution.
In our sample each event generated has associated a three-dimensional distribution of the energy density in the initial overlap region of the incident nuclei.
The energy density distribution is computed from the energy-momentum tensor given by the NeXuS code at an initial stage and can fluctuate on an event-by-event basis, depending on the nuclear distribution determined by the incident nuclei and by the collision impact parameter \cite{BrazJPhys.35.24}.
Alternatively, it is also possible to input a smooth distribution generated by averaging over many events.
For the analysis presented in the following we have divided the simulated sample into five event centrality classes for both fluctuating and smooth initial conditions, as described in table \ref{tab:CentralityClasses}.
\begin{table}[hbt]
\caption{Centrality classes of the NeXSPheRIO events used in the study.}
\begin{tabular}{c c c}
\hline\hline
Centrality                    &       \textit{b} range (fm) &              $\langle dN_{ch}/d\eta \rangle$ \\
\hline
0$-$10\%                      &              0.00 $-$ 4.78  &                              576.5 $\pm$ 0.4 \\
10$-$20\%                     &              4.78 $-$ 6.77  &                              393.8 $\pm$ 0.3 \\
20$-$30\%                     &              6.77 $-$ 8.29  &                              271.5 $\pm$ 0.2 \\
30$-$40\%                     &              8.29 $-$ 9.57  &                              182.3 $\pm$ 0.1 \\
40$-$60\%                     &              9.57 $-$ 11.72 &                               89.9 $\pm$ 0.1 \\
\hline\hline
\end{tabular}
\label{tab:CentralityClasses}
\end{table}
A minimum of 30 thousand events were used for each event centrality class, and for both smooth and fluctuating IC. The total simulated data sample adds to more than half million events.

\subsection{Flow comparisons}
\begin{figure}[ht]
\includegraphics[width=8.50cm]{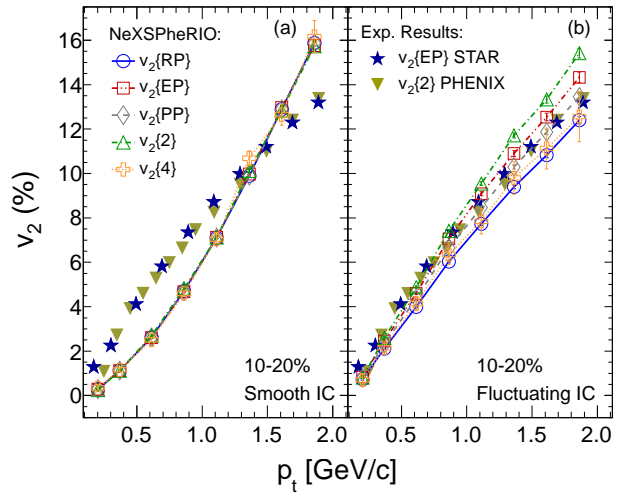}
\caption{\label{fig:v2vsPt_10to20} (color online) Differential $v_2$ as a function of the transverse momentum for 10-20\% central events. Open symbols are $v_2\{\mathrm{RP}\}$ (blue circles), $v_2\{\mathrm{EP}\}$ (red squares), $v_2\{\mathrm{PP}\}$ (grey diamonds), $v_2\{2\}$ (green triangles) and $v_2\{4\}$ (orange crosses), for smooth (panel (a)) and fluctuating (panel (b)) initial condition. Closed blue stars and yellow triangles are results from STAR \cite{PhysRevC.72.014904} and PHENIX \cite{PhysRevC.80.024909} experiments, respectively. The lines are just to guide the eyes.}
\end{figure}

\begin{figure}[ht]
\includegraphics[width=8.50cm]{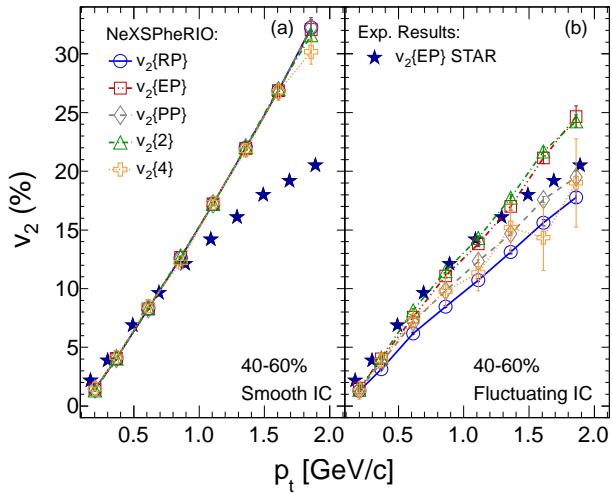}
\caption{\label{fig:v2vsPt_40to60} (color online) Differential $v_2$ as a function of the transverse momentum for 40-60\% peripheral events. Open symbols are $v_2\{\mathrm{RP}\}$ (blue circles), $v_2\{\mathrm{EP}\}$ (red squares), $v_2\{\mathrm{PP}\}$ (grey diamonds), $v_2\{2\}$ (green triangles) and $v_2\{4\}$ (orange crosses), for smooth (panel (a)) and fluctuating (panel (b)) initial condition. Closed blue stars are results from STAR \cite{PhysRevC.72.014904} experiment. The lines are just to guide the eyes.}
\end{figure}

\begin{figure}[ht]
\includegraphics[width=8.50cm]{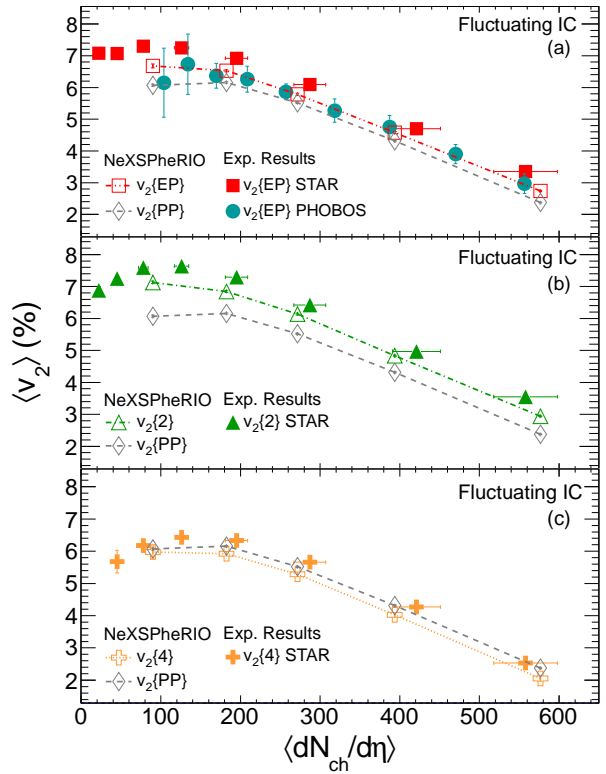}
\caption{\label{fig:v2vsdNchdetaExpData} (color online) Integrated $v_2$ as a function of $\langle dN_{ch} / d\eta \rangle$ at $\eta=0$. Panel (a) shows $v_2\{\mathrm{EP}\}$ for NeXSPheRIO events (red open squares) and results from STAR \cite{PhysRevC.72.014904} (red solid squares) and PHOBOS \cite{PhysRevC.72.051901} (green solid circles) experiments; panel (b) shows $v_2\{2\}$ for NeXSPheRIO events (green open triangles) and from STAR experiment (green solid triangles) \cite{PhysRevC.72.014904}; and panel (c) shows $v_2\{4\}$ for NeXSPheRIO events (orange open crosses) and from STAR experiment (orange solid crosses) \cite{PhysRevC.72.014904}. The $v_2$ results obtained with respect to the participant plane (grey open diamonds) were also included for reference. The lines are just to guide the eyes.}
\end{figure}

\begin{figure}[ht]
\includegraphics[width=8.50cm]{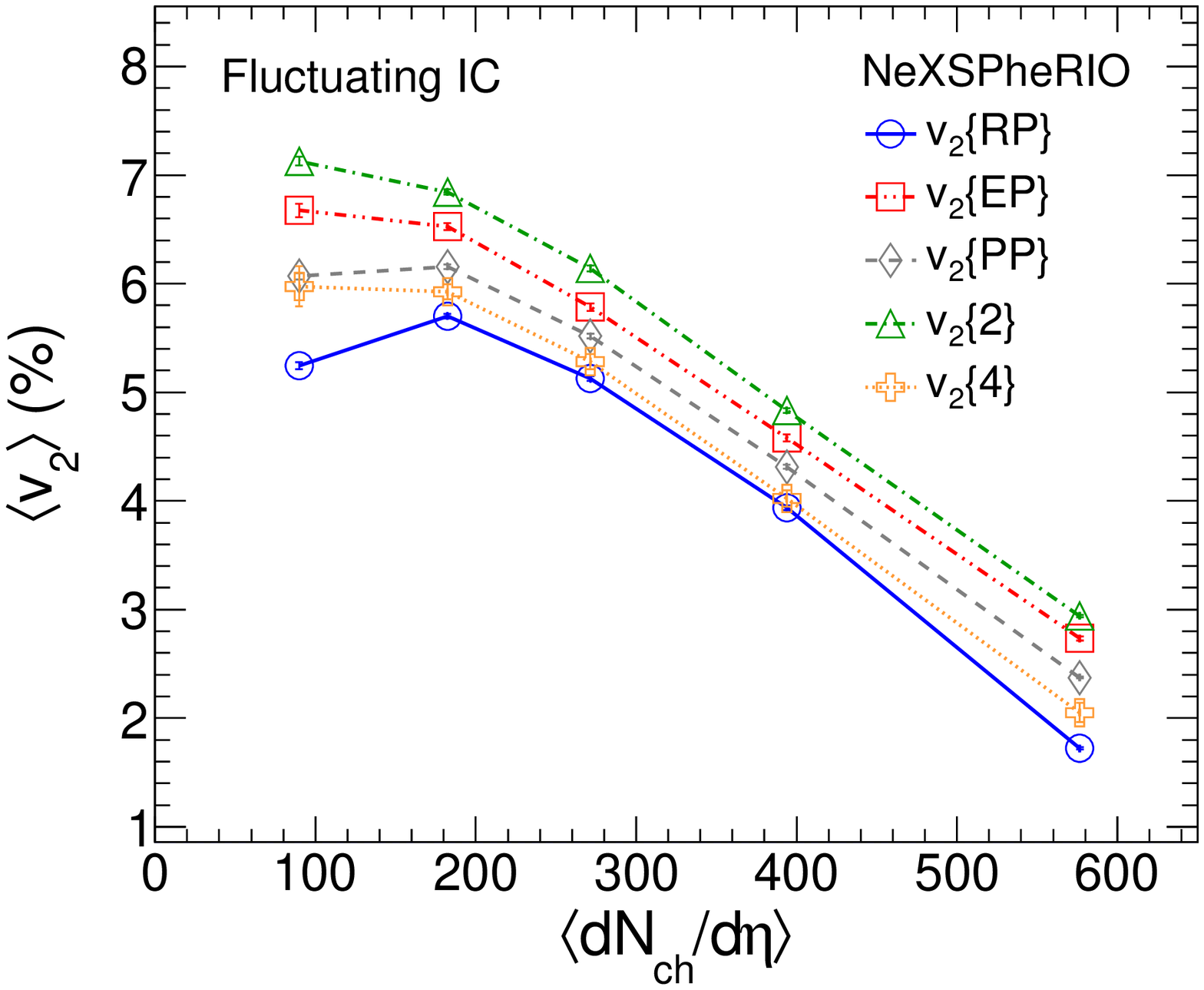}
\caption{\label{fig:v2vsdNchdeta} (color online) Mean integrated elliptic flow, $\langle v_2 \rangle$, as a function of $\langle dN_{ch} / d\eta \rangle$ at $\eta=0$. Blue open circles are $v_2\{\mathrm{RP}\}$, red open squares are $v_2\{\mathrm{EP}\}$, grey open diamonds are $v_2\{\mathrm{PP}\}$, green open triangles are $v_2\{2\}$ and orange open crosses are $v_2\{4\}$. The lines are just to guide the eyes.}
\end{figure}

\begin{figure}[ht]
\includegraphics[width=8.50cm]{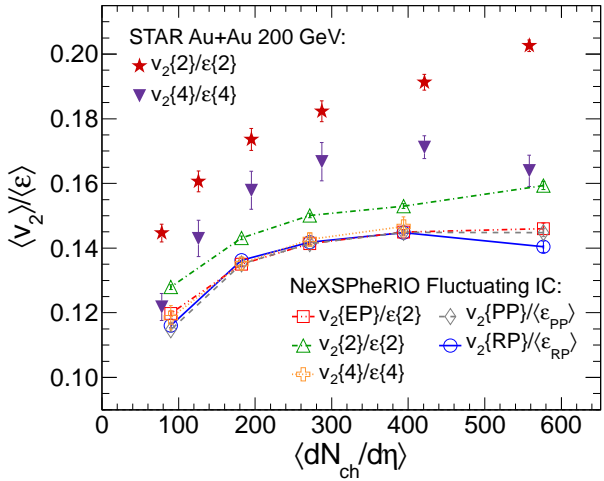}
\caption{\label{fig:v2OverEccentvsdNchdeta} (color online) Eccentricity scaled $v_2$ as a function of $\langle dN_{ch} / d\eta \rangle$ at $\eta=0$. Open symbols are $v_2\{\mathrm{PP}\} / \langle \varepsilon_{PP} \rangle$ (grey diamonds),  $v_2\{\mathrm{RP}\} / \langle \varepsilon_{RP} \rangle$ (blue circles), $v_2\{\mathrm{EP}\} / \varepsilon\{2\}$ (red squares), $v_2\{2\} / \varepsilon\{2\}$ (green triangles) and $v_2\{4\} / \varepsilon\{4\}$ (orange crosses), from NeXSPheRIO fluctuating IC.
Closed symbols are results from STAR experiment \cite{arXiv:1111.5637} for $v_2\{2\} / \varepsilon\{2\}$ (red stars) and $v_2\{4\} / \varepsilon\{4\}$ (blue triangles). The error bars are the quoted statistical and systematic uncertainties added in quadrature. The lines are just to guide the eyes.}
\end{figure}

We present results obtained for the $v_2$ estimates as a function of the transverse momentum, $p_t$, and centrality of the collision as given by the mean number of charged particles produced at $\eta=0$.
The calculations for all methods were performed using charged particles within the pseudorapidity window $|\eta| < 1.0$ and for $0.15 < p_t < 2.0$ GeV/c. 
The event plane determination was done using particles from $2.5<|\eta|<4.0$, therefore, avoiding auto-correlation on $v_2\{\mathrm{EP}\}$ calculation, and with a requirement of a minimum of 15 charged particles within this pseudorapidity region for each event.
We have also included published results reported by the STAR \cite{PhysRevC.72.014904}, PHENIX \cite{PhysRevC.80.024909} and PHOBOS \cite{PhysRevC.72.051901} experiments for comparison.
In the following plots (Figs. \ref{fig:v2vsPt_10to20} to \ref{fig:v2OverEccentvsdNchdeta}), the $v_2$ calculated by the different methods are presented by different symbols.
Blue open circles for the reaction plane $v_2$, red open squares for the $v_2$ calculated through the event plane method, gray open diamonds for the participant plane $v_2$, and the green open triangles and orange open crosses for the $v_2$ calculated using two and four particle cumulant methods, respectively.
In Figs. \ref{fig:v2vsPt_10to20} and \ref{fig:v2vsPt_40to60} we show the results for $v_2$ estimates as a function of transverse momentum for both smooth (left panels) and fluctuating (right panels) initial conditions, for centralities 10-20\% and 40-60\% respectively.
The blue star symbols are experimental results from the STAR experiment \cite{PhysRevC.72.014904} and represent $v_2$ estimates obtained with the event plane method, and the yellow triangles are results from the PHENIX experiment \cite{PhysRevC.80.024909} obtained with the second order cumulant method.
The $p_t$ dependence of the $v_2$ curve from the NeXSPheRIO data generated using fluctuating IC shows a better agreement to the experimental data than the results from the smooth IC.
The comparisons presented in Figs. \ref{fig:v2vsPt_10to20} and \ref{fig:v2vsPt_40to60} also show that in the case of the smooth IC, the different methods for the $v_2$ estimate that are based on the measurement of the final state particles provide the same results as the $v_2$ calculated from the IC participant plane method.
However, in the case of fluctuating IC, there is a discrepancy between the experimentally measurable $v_2$ methods and the actual elliptic flow from the participant plane calculation.
Moreover, the difference between the results from different $v_2$ methods increases with the transverse momentum.
This shows clearly how each method is affected by the fluctuations in the initial condition.
Assuming that the participant plane is the plane which defines the direction of the elliptic anisotropy created by the initial state and, therefore, defines the direction of the $v_2$ in the final state, it is possible to take the values of $v_2\{\mathrm{PP}\}$ as a reference for the real $\langle v_2 \rangle$.
Comparing to the results obtained with methods used experimentally (namely $v_2\{\mathrm{EP}\}$, $v_2\{2\}$ and $v_2\{4\}$), we see $v_2\{\mathrm{EP}\}$ above $v_2\{\mathrm{PP}\}$ and below $v_2\{2\}$, consistent with $\langle v_2 \rangle \le v_2\{\mathrm{EP}\} \le \sqrt{\langle v_2^2 \rangle}$ \cite{PhysRevC.77.014906}.

In Fig. \ref{fig:v2vsdNchdetaExpData} we show the dependence of the integrated $v_2$ with the collision centrality.
In the panel (a) we present the results obtained with the event plane method, panel (b) shows the results obtained with the second order cumulant, and panel (c) shows the results obtained with the fourth order cumulant.
Results from the STAR \cite{PhysRevC.72.014904} and PHOBOS \cite{PhysRevC.72.051901} experiments are also presented for comparison.
In all three panels we also included the participant plane $v_2$ as a reference.
The behavior of the flow obtained with the NeXSPheRIO model is very similar to the experimental results but the values are systematically lower.
In Fig. \ref{fig:v2vsdNchdeta} we have combined all the NeXSPheRIO results in the same panel and a clear ordering of the values from the different methods can be observed.
This ordering presented by $v_2\{4\}$, $v_2\{\mathrm{EP}\}$ and $v_2\{2\}$ is also observed in experimental results (see Ref. \cite{PhysRevC.72.014904}).
As expect, the elliptic flow calculated with respect to the reaction plane are lower than all other results, confirming that in a scenario with lumpy initial energy density distribution, the plane defined by the impact parameter vector and the beam axis no longer drives the preferred direction of the flow.
To see how the elliptic flow scales with the eccentricity of the initial state, we plotted the mean integrated $v_2$ obtained with the cumulant method over the respective eccentricity cumulant moment, calculated using the participant plane eccentricity \cite{arXiv:nucl-ex.0312008,PhysRevC.72.014904,PhysRevC.77.014906}:
\begin{eqnarray}
\varepsilon\{2\} &=& \sqrt{ \langle \varepsilon_{PP}^2 \rangle }, \label{eq:ecc2} \\
\varepsilon\{4\} &=& \sqrt[4]{ 2 \langle \varepsilon_{PP}^2 \rangle^2 - \langle \varepsilon_{PP}^4 \rangle }, \label{eq:ecc4}
\end{eqnarray}
as a function of the multiplicity density.
Comparing these ratios with $v_2\{\mathrm{RP}\} / \langle \varepsilon_{RP} \rangle$, $v_2\{\mathrm{PP}\} / \langle \varepsilon_{PP} \rangle$ and $v_2\{\mathrm{EP}\} / \varepsilon\{2\}$ as shown in Fig. \ref{fig:v2OverEccentvsdNchdeta}, obtained for $|\eta|<1.0$ (fluctuating IC), we find a good agreement among the NeXSPheRIO results. The points obtained with different methods seem to fall almost on top of each other, with the only exception of $v_2\{2\} / \varepsilon\{2\}$, which is systematically higher.
Such behavior is being investigated and we have already observed that the difference between $v_2\{2\} / \varepsilon\{2\}$ and the other ratios seems to vanish when increasing the pseudorapidity window considered in the calculations to $|\eta|<6.0$.
The observable $v_2\{4\}/\varepsilon\{4\}$ provides quite consistent results with $v_2\{\mathrm{PP}\}/\langle \varepsilon_{PP} \rangle$, the later one being a quantity obtained directly from the anisotropy of the IC and, therefore, not accessible experimentally.
For comparison, we also show results from STAR \cite{arXiv:1111.5637} experiment for $v_2\{2\} / \varepsilon\{2\}$ and $v_2\{4\} / \varepsilon\{4\}$, with the eccentricity taken from fKLN-CGC model, for Au+Au collisions at the center of mass energy of 200 GeV per nucleon pair.
The experimental results are always higher than the NeXSPheRIO results, but they present similar separation between $v_2\{2\} / \varepsilon\{2\}$ and $v_2\{4\} / \varepsilon\{4\}$.
Song {\it et al.} \cite{PhysRevLett.106.192301} have recently reported that the inclusion of viscous effects can reduce the baseline of ideal fluid $v_2/\varepsilon$. Moreover, they pointed out that a proper event-by-event treatment can also affect this baseline. Thus, a study of the observables in an event-by-event scenario, as presented here, is also important for the comparison between data and viscous hydrodynamical models.

\subsection{Flow fluctuations}
\begin{figure}[ht]
\includegraphics[width=8.50cm]{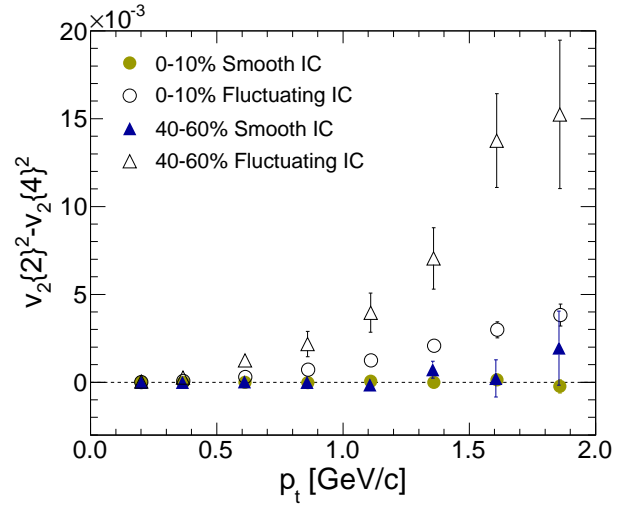}
\caption{\label{fig:absFluctVsPt} (color online) $v_2$ fluctuations as characterized by Eq. \ref{eq:absFluct} as a function of transverse momentum. Results from smooth IC are shown by closed symbols for 0-10\% central (yellow circles) and 40-60\% peripheral (blue triangles) events. Results from fluctuating IC are shown by open symbols for 0-10\% central (circles) and 40-60\% peripheral (triangles) events.}
\end{figure}

The differences observed for the flow estimates indicate that each method responds differently to the fluctuations in the initial condition, which can be used to study these fluctuations with the final observables.
In particular, from the definitions of flow estimates obtained with the cumulant formalism, it has been suggested by the authors of Ref. \cite{PhysRevC.80.014904} that, if only the leading order of $\sigma_{v_2}^2$ and $\delta$ are considered, then:
\begin{eqnarray}
v_{2}\{2\}^2 &\approx& \langle v_2 \rangle^2 + \delta + \sigma_{v_2}^2, \label{eq:v22Fluct} \\
v_{2}\{4\}^2 &\approx& \langle v_2 \rangle^2 - \sigma_{v_2}^2, \label{eq:v24Fluct}
\end{eqnarray}
where $\delta$ is the non-flow contribution (in general, correlations other than those related to the reaction plane), and $\sigma_{v_2}$ is the elliptic flow fluctuation.
The approximation in equation \ref{eq:v24Fluct} is valid for $\sigma_{v_2} \ll \langle v_2 \rangle$ and negligible higher order moments.
We note that this approximation breaks down for peripheral and central collisions where the skewness and kurtosis of the $v_2$ distribution and terms related to $\langle v_2 \rangle \sigma_{v_2}$ will contribute significantly to $v_2\{4\}$.
From equations \ref{eq:v22Fluct} and \ref{eq:v24Fluct}, we can extract that the difference between $v_2\{2\}^2$ and $v_2\{4\}^2$ will provide a quantitative measure of the non-flow contribution added to the flow fluctuation as given by: 
\begin{equation}
v_{2}\{2\}^2 - v_{2}\{4\}^2 \approx \delta + 2 \sigma_{v_2}^2.
\label{eq:absFluct}
\end{equation}
We present this quantity as a function of the transverse momentum in Fig. \ref{fig:absFluctVsPt}.
In order to reduce the statistical uncertainties in low multiplicity events, the particle pseudorapidity acceptance was increased to $|\eta|<6.0$.
Results from the smooth IC case, shown in Fig. \ref{fig:absFluctVsPt} as the solid yellow circles and the solid blue triangles, for event centrality 0-10\% and 40-60\% respectively, are consistent with negligible fluctuation and non-flow contributions in the model.
Results from the fluctuating IC events are presented in Fig. \ref{fig:absFluctVsPt} by open circles for the 0-10\% central events and open triangles for the 40-60\% peripheral events.
It is clear that in these cases, the quantity defined by equation \ref{eq:absFluct} is non-zero, indicating that indeed such parameter is sensitive to the fluctuations of the IC.
The absolute magnitude of the fluctuation is higher in the peripheral 40-60\% events compared to the central events.
\begin{figure}[ht]
\includegraphics[width=8.50cm]{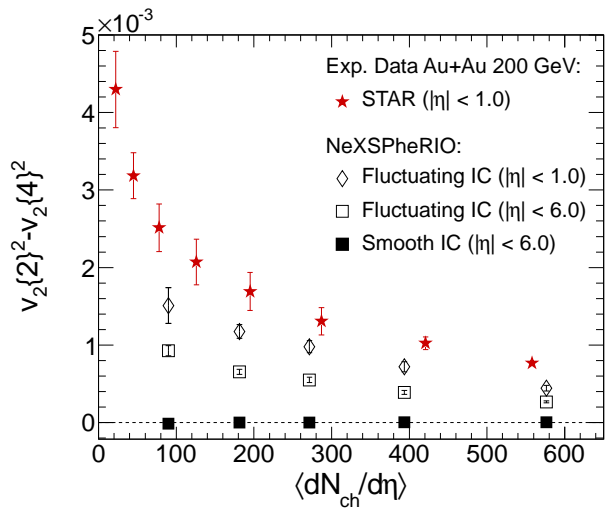}
\caption{\label{fig:absFluctVsdNchdeta} (color online) $v_2$ fluctuations as characterized by Eq. \ref{eq:absFluct} as a function of $\langle dN_{ch} / d\eta \rangle$ at $\eta=0$. The open diamonds are the results for the fluctuating IC case with $|\eta|<1.0$, and the squares are the results for the fluctuating IC (open squares) and smooth IC (solid squares), both for $|\eta|<6.0$. 
Red solid stars are results from STAR experiment \cite{arXiv:1111.5637}. The error bars are the quoted statistical and systematic uncertainties added in quadrature.}
\end{figure}
Fig. \ref{fig:absFluctVsdNchdeta} shows the results of $v_{2}\{2\}^2 - v_{2}\{4\}^2$ calculated for the different event centrality classes.
Solid squares represent the smooth IC case and open squares the fluctuating IC case, both for $|\eta| < 6.0$.
The smooth IC case is consistent with zero, while the fluctuating IC case shows a steady decrease towards the most central collisions.
In addition, we have also performed the analysis for fluctuating IC obtained within a tighter pseudorapidity window ($|\eta|<1.0$), represented by open diamonds in Fig. \ref{fig:absFluctVsdNchdeta}.
We observe a similar behavior compared to the fluctuating IC case for wider $\eta$ window, but with points systematically higher.
The increase of the values observed for this case suggests a dependence of the fluctuations with pseudorapidity window.
Published data from STAR experiment \cite{arXiv:1111.5637} are also shown for $|\eta|<1.0$.
These points present values always higher than the results from NeXSPheRIO, but the behavior is similar.
Moreover, the contribution of non-flow effects is expected to be greater in experimental data.

\begin{figure}[ht]
\includegraphics[width=8.50cm]{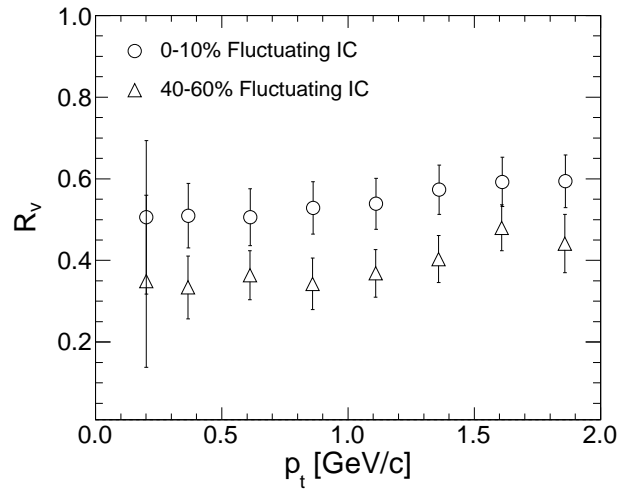}
\caption{\label{fig:relFluctVsPt} $R_v$ parameter as a function of the transverse momentum. Open circles represent the 0-10\% most central events and open triangles are for 40-60\% peripheral events.}
\end{figure}

\begin{figure}[ht]
\includegraphics[width=8.50cm]{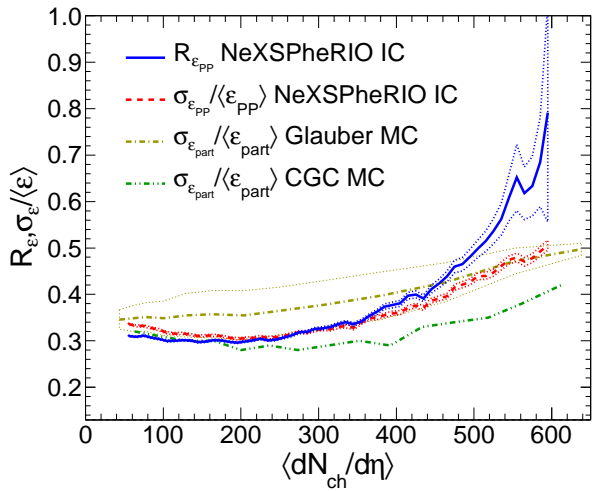}
\caption{\label{fig:eccRelFluctVsdNchdeta} (color online) Comparison between the $R_{\varepsilon}$ parameter (blue solid line) and the eccentricity relative fluctuation, $\sigma_{\varepsilon}/\langle \varepsilon \rangle$ (red dashed line) calculated with respect to the participant plane in the NeXSPheRIO initial condition, as a function of $\langle dN_{ch} / d\eta \rangle$ at $\eta=0$. It is also shown the relative eccentricity fluctuation calculated with Glauber Model (yellow dot-dashed line) and CGC (green double-dot-dashed line) Monte Carlo \cite{PhysRevC.76.041903,*PhysRevLett.104.142301}.}
\end{figure}

Assuming a negligible non-flow component in equation \ref{eq:absFluct}, it is possible to take the difference $v_{2}\{2\}^2 - v_{2}\{4\}^2$ as an estimate of the absolute flow fluctuation.
An estimate for the relative flow fluctuation can also be defined by using the equations \ref{eq:v22Fluct} and \ref{eq:v24Fluct} as:
\begin{equation}
R_v = \sqrt{\frac{v_{2}\{2\}^2 - v_{2}\{4\}^2}{v_{2}\{2\}^2 + v_{2}\{4\}^2}}.
\label{eq:Rv}
\end{equation}
In the case that the $v_2$ distribution is Gaussian-like, with the mean much larger than the width, the quantity $R_v$ is a resonable approximation for relative flow fluctuation $\sigma_{v_2}/\langle v_2 \rangle$ \cite{PhysLett.B659.537,arXiv:0905.0174}.
In Fig. \ref{fig:relFluctVsPt} the $R_v$ parameter as a function of $p_t$ is presented for the most central 0-10\% and the peripheral 40-60\% event centrality classes. 
Opposite to what was observed for the absolute flow fluctuations in Fig. \ref{fig:absFluctVsPt}, in this case the central events show higher values than the peripheral events.
The higher value of $R_v$ observed in central events compared to the peripheral are due to both the increase of the fluctuations in central collisions and also the increase of asymmetry in the $v_2$ distributions.
The dependence of $R_v$ with centrality will be further discussed next.
In addition, we observe negligible dependence with transverse momentum up to 2 GeV/c for both central and peripheral events.
Similarly to equation \ref{eq:Rv}, it is possible to define an equivalent quantity for the eccentricity, $R_{\varepsilon}$:
\begin{equation}
R_{\varepsilon} = \sqrt{\frac{\varepsilon\{2\}^2 - \varepsilon\{4\}^2}{\varepsilon\{2\}^2 + \varepsilon\{4\}^2}},
\label{eq:Re}
\end{equation}
where $\varepsilon\{2\}$ and $\varepsilon\{4\}$ are given by the equations \ref{eq:ecc2} and \ref{eq:ecc4}, respectively.
Since we have access in our simulated events to both the IC and also the final state particles, we can verify if the quantity $R_{\varepsilon}$ is a good approximation for the relative eccentricity fluctuation, $\sigma_{\varepsilon}/\langle \varepsilon \rangle$.
In Fig. \ref{fig:eccRelFluctVsdNchdeta} we show the $R_{\varepsilon}$ parameter as defined by equation \ref{eq:Re} (blue solid line) and the relative eccentricity fluctuation, $\sigma_{\varepsilon}/\langle \varepsilon \rangle$ (red dashed line), extracted from the participant plane eccentricity distribution of Fig. \ref{fig:EccentVsdNchdeta}, panel (b).
We have also included for comparison, results for Glauber Model (yellow dot-dashed line) and Color Glass Condensate (CGC) (green double-dot-dashed line) Monte Carlo calculations \cite{PhysRevC.76.041903,*PhysRevLett.104.142301}. In this case, the references reported these curves as a function of the number of participants in the collision ($N_{part}$). We used the values presented in table II of Ref. \cite{PhysRevC.79.034909} to convert $N_{part}$ to the equivalent $\langle dN_{ch}/d\eta \rangle$ at $\eta=0$, used in this work to summarize the collision centrality.
This result shows that the observable $R_v$ is a good measure of the relative fluctuations in mid-central collisions, but overestimates the fluctuations in central collisions. 
The higher value of $R_v$ in central events is mainly due to the contributions of higher moments of the $\varepsilon_{PP}$ distribution such as skewness and kurtosis. 
The Gaussinan-like condition is no longer satisfied and the approximations done in equation \ref{eq:v24Fluct} that allows for the interpretation that $R_v$ is a good measure of the flow fluctuation is no longer valid.

As an extension of the work already presented by Hama and collaborators \cite{arXiv:0711.4544} and Sorensen \cite{JPhysG.34.S897}, we have also plotted the $R_v$ parameter as given by the equation \ref{eq:Rv} as a function of the mean number of charged particles produced at $\eta=0$, hence, calculated for the different event centrality classes.
\begin{figure}[ht]
\includegraphics[width=8.50cm]{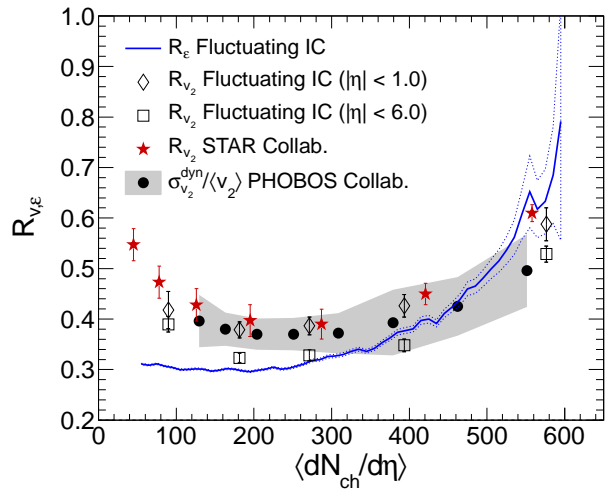}
\caption{\label{fig:relFluctVsdNchdeta} (color online) Comparison between the $R_v$ and $R_{\varepsilon}$ parameters as a function of $\langle dN_{ch}/d\eta \rangle$ at $\eta=0$. The symbols represent $R_v$ for the fluctuating IC case with $|\eta|<6.0$ (open squares) and $|\eta|<1.0$ (open diamonds). The blue solid line is the $R_{\varepsilon}$ parameter calculated with respect to the participant plane in the NeXSPheRIO IC. Closed symbols are published results from PHOBOS \cite{PhysRevLett.104.142301} (black circles) and STAR \cite{arXiv:1111.5637} (red stars) experiments. The shaded band represent the errors quoted by PHOBOS and the error bars in STAR data are the statistical and systematic uncertainties added in quadrature.}

\end{figure}
The results are presented in Fig. \ref{fig:relFluctVsdNchdeta}.
Open diamonds were obtained for $|\eta| < 1.0$ and the open squares for $|\eta| < 6.0$.
In addition to the $R_v$ parameter we have also included the $R_{\varepsilon}$ parameter (blue solid line) calculated with respect to the participant plane obtained from the IC.
Also, results obtained for the dynamic relative flow fluctuation, $\sigma_{v_2}^{dyn} / \langle v_2 \rangle$ (black closed circles), reported by the PHOBOS experiment \cite{PhysRevLett.104.142301}, and the $R_{v}$ parameter (red closed stars) reported by the STAR experiment \cite{arXiv:1111.5637}, are shown for comparison.
Therefore, since the final state observable that we have is $R_{v}$, it is important to compare it with the initial state quantity $R_{\varepsilon}$, and not $\sigma_{\varepsilon}/\langle\varepsilon\rangle$.
The NeXSPheRIO values of $R_v$ for $|\eta|<1.0$ seem to agree very well with the experimental results.
Even though a perfect agreement between the $R_{\varepsilon}$ curve and the points calculated with the $R_v$ parameter is not observed, it is remarkable that those quantities, obtained using the properties of the very beginning and the very final stage of the system evolution, still yield such similar values.
Thus, we can conclude that the experimentally observable parameter $R_v$ is a good estimate of the fluctuations in the initial state of the collisions, for mid-central events, which cannot be probed directly.
Moreover, we can conclude that the $R_v$ parameter does not get affected by the hydrodynamic evolution and the freeze-out process.

\section{Summary} \label{sec:summary}
In this work we have analyzed a large amount of simulated events produced with the NeXSPheRIO code and both the IC and the final state particles were saved and analyzed on an event-by-event basis.
Smooth and fluctuating IC were used to generate the events for five different centrality classes.
This allowed us to study the effects of fluctuations in the initial energy density distribution through the $v_2$ estimates from the final particle azimuthal distributions.
We showed for the smooth IC cases that the methods generally used in experimental data analysis to estimate $v_2$, namely $v_2\{\mathrm{EP}\}$, $v_2\{\mathrm{2}\}$ and $v_2\{\mathrm{4}\}$, produce consistent results with flow estimates that uses informations from the initial condition, namely $v_2\{\mathrm{RP}\}$ and $v_2\{\mathrm{PP}\}$.
Also, for the fluctuating IC we found that the methods start to deviate from each other when going to higher transverse momentum, giving rise to a systematic ordering as a function of centrality.
The discrepancy between the different $v_2$ calculation methods were exploited to be used as a measurement of the IC degree of fluctuation.
Our results show that the magnitude of the non-flow plus the flow fluctuations increases with $p_t$ for both central and peripheral events, being more pronounced for the later one.
On the other hand, the relative fluctuations, as defined by the $R_v$ parameter, shows no dependence with transverse momentum.
We also observed a dependence of the quantity defined in equation \ref{eq:absFluct} with the collision centrality and the pseudo-rapidity window of the particles used in the analysis.
Although the behavior observed for $R_v$ and $R_{\varepsilon}$ as a function of the mean number of charged particles at $\eta=0$ do not completely agree, the values obtained are remarkably similar, which indicates that indeed the measurements of the final state flow fluctuations may provide important information on the initial state fluctuations.

\section*{Acknowledgments}
This work was supported in part by FAPESP, FAPERJ, CNPq, CAPES, and PRONEX of Brazil, and by the Offices of NP and HEP within the U.S. DOE Office of Science under the contracts of DE-FG02-88ER40412 and DE-AC02-98CH10886.

%

\end{document}